\documentclass[a4paper,fleqn,usenatbib,useAMS]{mnras}

\usepackage{graphicx}	
\usepackage{grffile}
\usepackage{amsmath}	
\usepackage{amssymb}	
\usepackage{multicol}        

\usepackage{bm}		
\usepackage{pdflscape}	
\usepackage[autostyle]{csquotes}
\usepackage{caption}
\usepackage[export]{adjustbox}
\usepackage{endnotes}
\usepackage{subcaption}
\usepackage{floatrow}
\usepackage{hyperref}

\usepackage[T1]{fontenc}
\usepackage{ae,aecompl}

\usepackage{newtxtext,newtxmath}

\title[\textit{On the weak magnetic field of millisecond pulsars}]{{ On the weak magnetic field of millisecond pulsars: Does it decay \emph{before} accretion?}}

\author[Cruces, Reisenegger, \& Tauris]{Marilyn Cruces$^{1,2}$ \thanks{Contact e-mail: \href{mscruces@uc.cl}{mscruces@uc.cl}}, Andreas Reisenegger$^{1}$ and Thomas M. Tauris$^{3,4,2}$ 
\\
$^{1}$Instituto de Astrof{\'\i}sica, Facultad de F{\'\i}sica, Pontificia Universidad Cat\'olica de Chile, Av. Vicu\~na Mackenna 4860, Macul 7820436, Santiago, Chile
\\
$^{2}$Max-Planck-Institut f{\"u}r Radioastronomie, Auf dem H{\"u}gel 69, D-53121 Bonn, Germany
\\
$^{3}$Aarhus Institute of Advanced Studies (AIAS), Aarhus University, H{\o}egh-Guldbergs~Gade~6B, 8000~Aarhus~C, Denmark
\\
$^{4}$Department of Physics and Astronomy, Aarhus University, Ny Munkegade 120, 8000~Aarhus~C, Denmark
}

\date{\today}

\pubyear{2018}

\begin{document}
\label{firstpage}
\pagerange{\pageref{firstpage}--\pageref{lastpage}}
\maketitle

\begin{abstract}
Millisecond pulsars are old, fast spinning neutron stars thought to have evolved from classical pulsars in binary systems, where the rapid rotation is caused by the accretion of matter and angular momentum from their companion. During this
transition between classical and millisecond pulsars, there is a
magnetic field reduction
of $\sim 4$ orders of magnitude, which is not well understood. According to the standard scenario, the magnetic field is reduced as a consequence of accretion, either through ohmic dissipation or through screening by the accreted matter. We explored an alternative hypothesis in which the magnetic field is reduced through \textit{ambipolar diffusion} before the accretion. This is particularly effective during the long epoch in which the pulsar has cooled, but has not yet started accreting. This makes the final magnetic field dependent on the evolution time of the companion star and thus its initial mass. We use observed binary systems to constrain the time available for the magnetic field decay based on the current pulsar companion: a helium white dwarf, a carbon-oxygen white dwarf, or another neutron star. Based on a simplified model without baryon pairing, we show that the proposed process agrees with the general distribution of observed magnetic field strengths in binaries, but is not able to explain some mildly recycled pulsars where no significant decay appears to have occurred. We discuss the possibility of other formation channels for these systems and the conditions under which the magnetic field evolution would be set by the neutron star crust rather than the core.
\end{abstract}




\begingroup
\let\clearpage\relax
\endgroup
\section{Introduction}

Millisecond pulsars (MSPs) are old neutron stars (NSs) thought to have evolved from classical pulsars in binary systems, where they have obtained their fast rotation through the accretion of matter and angular momentum from a companion star in a close binary system \citep{Alpar82,Bhattacharya91,Tauris06}. MSPs also have substantially weaker surface dipole magnetic fields compared to the normal population of young radio pulsars, typically $B\sim 10^{8-9}$ G vs. $10^{11-13}$ G, respectively \citep{Manchester05}, which is usually also explained as a result of accretion \citep{Bisnovatyi-Kogan74,Bhattacharya11}, a hypothesis we will hereafter call ``accretion scenario''. One mechanism proposed for this is the enhanced resistivity of the NS crust as it is heated by the accretion \citep{Geppert94}. However, this would require the magnetic flux to go exclusively through the crust of the NS, which is implausible, unless it is somehow expelled from the core, e.g. through a strong Meissner effect or a very high resistivity. Another proposal is screening of the magnetic field by the accreted matter \citep{Romani90}, which in turn has the difficulty of magneto-hydrodynamic (MHD) instabilities \citep{Mukherjee13a, Mukherjee13b}.\\ 
\\
In this paper, we explore the alternative hypothesis of a magnetic field decay produced before the accretion process through  \textit{ambipolar diffusion} \citep{Goldreich92}, in which the magnetic flux is transported by the charged particles (protons and electrons) in the NS core as these are pushed by the Lorentz force, making them move relative to the neutrons. Since the collision rate between charged particles and neutrons is strongly dependent on temperature, this mechanism (``diffusion scenario'') becomes effective at late times ($\gtrsim 10^6\mathrm{yr}$) for non-superfluid cores, once the NS has cooled down, but before it is reheated by accretion.\\ 
\\
We do not attempt to construct a realistic, quantitatively reliable model, which would be highly complex, requiring 3-dimensional multi-fluid MHD simulations with largely unknown ingredients such as the impurity content of the NS crust and the mutual interactions of superfluid, superconducting, and normal particles of different kinds in the core. Instead, we present a ``toy model'' for the coupled evolution of a NS magnetic field, temperature, and rotation, with each of these variables represented by a single number and ignoring the presence of possible superfluid and superconducting states and a solid crust. The purpose of this model is to explore the plausibility of the diffusion 
hypothesis by comparing the results of this model with the observed magnetic fields of pulsars with degenerate companions, in which the main-sequence lifetime of the companion's progenitor star determines the time available for this process. A more detailed model must be left for future work.\\
\\
This paper is organized as follows. In section \ref{sec:amb.diff}, we describe the mechanism of \textit{ambipolar diffusion}. In section \ref{sec:mod}, we show the model used for the evolution of the NS's core temperature, rotation, and magnetic field strength. In section \ref{sec:Results}, we present the results on how this evolution proceeds and the predictions on how a NS will move on the $P-\dot P$ plane as first ambipolar diffusion reduces the magnetic field and then accretion increases the rotation rate. In section \ref{sec:Discussion}, we discuss the implications of our results, and the shortcomings of our model, while in section \ref{sec:Conclusions}, we present the conclusions of our work.

\section{Model}
\label{sec:mod}
\subsection{Ambipolar diffusion in the NS core}
\label{sec:amb.diff}

NS magnetic fields probably have their origin in the field of their non-degenerate progenitor stars, which is amplified by flux freezing during the collapse and further by differential rotation, instabilities, and convective motions in the proto-NS following the supernova explosion. Once these motions have settled, the magnetic field reaches an equilibrium configuration that is likely to fill the whole volume of the NS, stabilized by the composition gradient (radially decreasing proton/neutron ratio) in the NS core \citep{Reisenegger92,Goldreich92,Reisenegger09}. Since the conductivity is very high \citep{Baym69}, the magnetic field is effectively frozen into the charged particles (e.~g., \citealt{Spruit13}; \citealt{Thorne17}). Thus, for the magnetic field to evolve, the charged particles have to move, somehow overcoming the composition gradient. This can happen in two different ways, which are effective in opposite temperature regimes (\citealt{Goldreich92}; see also \citealt{Hoyos08,Hoyos10}):
\begin{itemize}
\item[(i)] At \textit{high temperatures}, charged particles and neutrons are strongly coupled to each other by collisions, but they can convert into each other by weak interactions (\emph{Urca reactions}). Thus, the core matter can be considered as a single fluid that gradually changes its composition as it moves radially at a rate proportional to the Urca reaction rate, but much slower than the neutrino cooling time (\citealt{Reisenegger09}; \citealt{Ofengeim18}). This is likely to be important in the case of magnetars, where the heat generated by the dissipation of the magnetic energy might keep the NS core hot enough so a substantial decay can actually happen \citep{Thompson96,Reisenegger09}. 
\item[(ii)] At \textit{low temperatures}, Urca reactions are essentially frozen, but the collision rate between neutrons and charged particles is also strongly suppressed, therefore it becomes possible for these two components to move separately, with different velocity fields, in this way allowing the composition to adjust to its equilibrium state at any given density. This relative motion between neutrons and charged particles, known as ``ambipolar diffusion'', is the focus of the present work. 
\end{itemize}
In the simplified model of \citet{Goldreich92}, which takes the neutrons as a fixed background and ignores the possible presence of superfluids or superconductors, the magnetic field $\vec B$ is transported by the charged particles at the ambipolar diffusion velocity
\begin{equation}\label{Amb_velocity}
\vec v_{AD}= \frac{\vec f_B^s}{n_c\left(m_p/\tau_{pn}+m_e^*/\tau_{en}\right)},
\end{equation}
where $\vec f_B^s$ is the solenoidal part of the magnetic force density (see \citealt{Goldreich92} for details), $n_c$ is the number density of charged particles, $m_p$ is the proton mass, $m_e*$ is the effective electron mass, and  $\tau_{ij}$ is the mean time between collisions of particles of species $i$ against species $j$. Together with the advection equation for the magnetic field,
\begin{equation}\label{B_Evolution}
\frac{\partial \vec B}{\partial t}=\nabla\times\left(\vec v_{AD} \times \vec B\right)
\end{equation}
this yields the characteristic time scale for ambipolar diffusion,
\begin{equation}\label{amb_timescale}
t_{AD}= 3\times 10^9\,\frac{T_4^2L_5^2}{B_8^2}\,\rm{yr},
\end{equation}
\citep{Goldreich92}, here normalized in convenient units whose relevance will be clear in section \ref{sec:Results}: $T_4$ is the core temperature in units of $10^4$ K, $B_8=B/(10^8\,\mathrm{G})$, and $L_5$ is the characteristic lengthscale of magnetic field gradients in units of $10^5$ cm. The typical decay time obtained, $\sim\mathrm{Gyr}$, is of the order of expected evolution times of the low-mass progenitors of white dwarfs (WDs), and might thus be probed in NS-WD binary systems, where the WD acts as a clock.


\subsection{The crust as an effective vacuum}

In order to affect the surface dipole field inferred from the pulsar spin-down rate, the processes happening in the NS core must somehow be transmitted through the solid crust. If the Lorentz forces are strong enough to break the crust, they could move the crustal matter and in this way rearrange the surface field. This is plausible in the case of magnetars, but probably not in rotation-powered pulsars, much less once their magnetic field has been reduced close to MSP levels. Another possibility is a crust with a resistive (Ohmic) diffusion time shorter than the ambipolar diffusion time in the core, which will also allow a rapid rearrangement of the surface field controlled by the processes going on in the core. In order to maintain a high resistivity even at low temperatures, the crustal solid must have a high impurity content.\\ 
\\
The timescale for Ohmic diffusion due to impurity scattering is \citep{Cumming04}
\begin{equation}\label{Ohm_timescale}
t_{Ohm}= 5.7\,\mathrm{Myr} \frac{\rho_{14}^{5/3}}{Q}\left(\frac{Z}{30}\right)\left(\frac{Y_e}{0.05}\right)^{1/3}\left(\frac{Y_n}{0.8}\right)^{10/3}\left(\frac{f}{0.5}\right)^2\left(\frac{g_{14}}{2.45}\right)^{-2},
\end{equation}
where $\rho_{14}$ is the crust density in units of $10^{14}\mathrm{g\,cm}^{-3}$, $Z$ is the atomic number of the dominant nuclei, $Y_e$ and $Y_n$ are the electron and neutron fraction, respectively, and $f$ is a factor that accounts for interactions between neutrons. By far the most uncertain variable is the \textit{impurity parameter} $Q$, which quantifies the root-mean-square deviations from the crust average composition. Its value has been theoretically estimated to lie anywhere between $10^{-3}$ \citep{Flowers77} for very pure crusts that have not been subject to accretion up to 100 for accreted crusts where the accreted material has replaced the original crust \citep{Schatz99}. Recent models for the observed thermal relaxation of transient accretors after outbursts require a relatively low impurity parameter ($Q\sim 1$) at low densities ($\rho<8\times 10^{13}\,\mathrm{g\,cm}^{-3}$) and a substantially higher value ($Q\sim 20$) in the deepest and densest layers \citep{Deibel17}. 
Although it is not clear how representative these values are for a crust that has \emph{not} undergone accretion, this makes it plausible to have $t_{Ohm}\ll t_{AD}$. In this limit, essentially no currents flow in the crust, so the latter will act as an extension of the near-vacuum outside the star, and the surface magnetic field will be determined by the bottleneck in the core.

\subsection{Magneto-thermo-roto-chemical evolution in the core}

When analyzing the magnetic field decay in old NSs, we must take into account that the decay rate depends on the core temperature $T$, which itself is evolving. At early times, $T$ decreases mainly through the emission of neutrinos produced by Urca reactions in the core, and later through the emission of thermal photons from the surface. If no reheating mechanisms were present, the temperature would decay to extremely low values within $\sim 10^7\,\mathrm{yr}$. However, several reheating mechanisms have been proposed in the literature (see \citealt{Gonzalez10} for a summary and references), and the detection of likely thermal ultraviolet emission from three pulsars in the age range $\sim 10^{7-10}\mathrm{yr}$ \citep{Kargaltsev04,Durant12,Rangelov17,Pavlov17}, indicating surface temperatures $T_s\sim 10^5\mathrm{K}$, appears to confirm this prediction. Among the mechanisms proposed, \citet{Gonzalez10} found  that two of them are most promising to explain the thermal emission of very old, low-B NSs. One of these is the friction caused by the motion of superfluid neutron vortices in the NS crust \citep{Alpar84}, which depends on the very uncertain angular momentum excess $J$  in the crustal superfluid. The other is ``rotochemical heating'': As the NS rotation slows down, the star contracts, causing chemical imbalances among the particle species present, e.~g.,
\begin{equation}\label{imbalance}
\eta_{npe}\equiv\mu_n-\mu_p-\mu_e > 0 \quad\mathrm{and}\quad \eta_{np\mu}\equiv\mu_n-\mu_p-\mu_\mu > 0,
\end{equation}
where $\mu_i$ is the chemical potential (roughly the Fermi energy) of particle species $i$, and the labels $n$, $p$, $e$, and $\mu$ stand for neutrons, protons, electrons, and muons, although other particle species might also be present and involved in these processes. This induces non-equilibrium Urca reactions that deposit energy inside the NS, keeping it warm for as long as it keeps spinning down \citep{Reisenegger95,Fernandez05}. Similarly, various spin-down-induced nuclear reactions in the NS crust (e.~g., \citealt{Haensel90}) also reheat the star, but only in the case of NSs whose crust has previously been compressed by a substantial amount of accretion \citep{Gusakov15}.\\
\\
Here, we use the code of \citet{Petrovich10} for the coupled evolution of the temperature and the chemical imbalances with neutrino and photon cooling as well as rotochemical heating, considering only modified Urca reactions without Cooper pairing gaps. To this, we add evolution equations for the magnetic field undergoing ambipolar diffusion,
\begin{equation}
\dot{B}\approx-\frac{B}{t_{AD}}\label{B-evo},
\end{equation}
and the decreasing rotation rate due to magnetic dipole spin-down,
\begin{equation}\label{dipole}
\dot\Omega=-\frac{2R_c^6}{3c^3I}B^2\Omega^3
\end{equation}
where $R_c$ and $I$ are the star's core radius and moment of inertia, and $c$ is the speed of light, as well as a magnetic dissipation term in the thermal evolution equation, with total power
\begin{equation}
\dot E_B\approx\frac{1}{3} R_c^3B\dot{B},\label{B-energy}
\end{equation}
We note that the magnetic field evolution is treated very schematically, with a single scalar variable $B$ representing a potentially complex vector field, and assuming that it decays on the ambipolar diffusion timescale, very similar to the approach of \citet{Xia13}. This assumption is contradicted by simulations of ambipolar diffusion in axial symmetry, which show that the magnetic field relaxes to a stable equilibrium state in which the Lorentz force is balanced by a pressure gradient in the charged particles \citep{Castillo17}, so the dissipation stops. However, non-axisymmetric instabilities likely lead to a full decay, as observed in 3-dimensional MHD simulations \citep{Mitchell15}.\\
\\
Our code allows us to track the core temperature $T_c$, surface temperature $T_s$, chemical imbalances $\eta_{npe}$ and $\eta_{np\mu}$, magnetic field strength $B$, spin period $P$ and its derivative $\dot{P}$ as functions of time. The values of these outputs are governed by the initial spin-period $P_0$, the initial field strength $B_0$, the lengthscale $L$ of spatial variations of the magnetic field, and the time available for the evolution. 

\subsection{Constraining the time available for magnetic field decay}
\label{sec:constrainingT}

In the proposed scenario, the magnetic field of NSs in binary systems will decay until their companion star initiates transferring mass onto their surface, increasing their core temperature to $\sim 10^8\mathrm{K}$ and thus choking the ambipolar diffusion. After accretion stops, internal reheating processes in the much faster rotating NS are expected to keep the core temperature high enough for ambipolar diffusion to remain negligible. This prevents further decay of the surface B-field below residual values of the order $\sim 10^8\;{\rm G}$, as observed in recycled radio MSPs with old WD companions. 
As a rough estimate, we assume that the accretion starts once the companion star ends its main-sequence lifetime, which thus sets the time available for magnetic field decay in the NS as: 
\begin{equation}\label{t_{MS}}
t_{MS}\approx 10^{10}\mathrm{yr} \left(\frac{M_{MS}}{M_\odot}\right)^{-2.5}
\end{equation}
\citep{kw90}, which roughly holds for main sequence masses, $M_{MS}$ in the range $0.1-50\,M_\odot$. Of course this ignores the time needed for the formation of the pulsar. In the case of a WD companion, the evolutionary time of the (more massive) NS progenitor is much shorter than that of the (less massive) WD progenitor, so the correction is negligible. This may not hold for double NS systems, where the progenitor masses might have been similar, thus leading to similar and unknown evolutionary times. In this latter case, the already very short time estimated in our approach might still be an overestimate of the actual time available. Finally, we note that for short orbital period MSPs (less than a few days), the assumption of a pre-accretion NS B-field decay timescale equal to the main-sequence lifetime of its companion star progenitor is somewhat an overestimate as the accretion is often a result of Case A Roche-lobe overflow (RLO) in low-mass X-ray binaries (LMXBs), whereas it is a slight underestimate in wider orbit MSPs where the companion star is a red giant prior to RLO \citep{Tauris99}. In any case, equation~(\ref{t_{MS}}) is a reasonable approximation, within a factor of two, for the timescale in which ambipolar diffusion is active in the NS core.
There are currently 286 binary radio pulsars known, of which compact object companion stars are found to include: He~WDs (121), CO~WDs (40) and NSs (19), according to the latest version 1.59 of the ATNF Pulsar Catalogue \citep{Manchester05}. Typical examples of main-sequence masses of their progenitor stars \citep[e.g.][]{Tauris11} are shown in Table $\ref{table_ta}$.
In the rest of this work, we disregard the 78 binary pulsars which are found in globular clusters, since their evolutionary history is uncertain as it might involve encounter events in such a dense stellar environment whereby the companion star is exchanged and information of the binary origin is lost.
\begin{table} 
\caption{Companion types, lower and upper boundaries of the main sequence progenitor mass range that can be confidently assumed to produce this kind of remnant \citep{Tauris12}, and respective main-sequence lifetime estimated from equation~(\ref{t_{MS}}).
\label{table_ta}}
\begin{tabular}{lrc}
 \hline
  Companion & $M_{pc}$ & $t_{MS}$\\
  type & [$M_{\sun}$] & [yr]\\
  \hline
  \hline
  helium WD (He~WD) & 1 & $1\times 10^{10}$\\
  &1.6 & $3\times 10^{9}$\\
  massive WD (CO~WD) & 3 & $6\times 10^{8}$\\
  & 6 & $1\times 10^{8}$\\
  neutron star (NS) & 10 & $3\times 10^{7}$\\
  & 25 & $3\times 10^{6}$\\
  \hline
\end{tabular}

\end{table}

\section{Results}
\label{sec:Results}
\subsection{Coupled evolution of physical variables}

\begin{figure}
\vspace{-0.5cm}
\begin{subfigure}[b]{1\textwidth}
	\caption{Spin period evolution}
	 \includegraphics[width=1\textwidth]{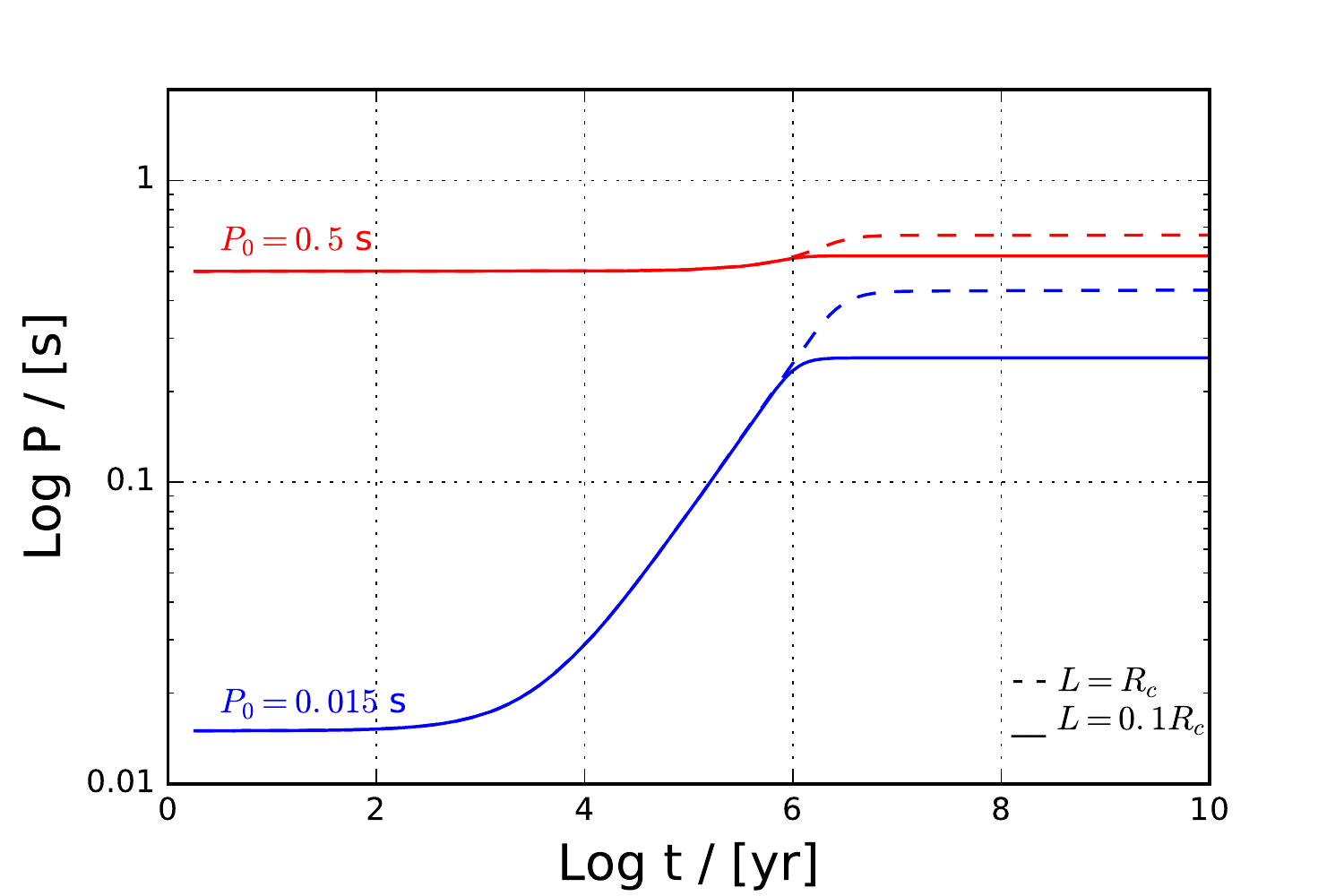} 
	\label{P-evol}
	\end{subfigure}

\begin{subfigure}[b]{1\textwidth}
	\caption{Magnetic field strength evolution}
	\includegraphics[width=1\textwidth]{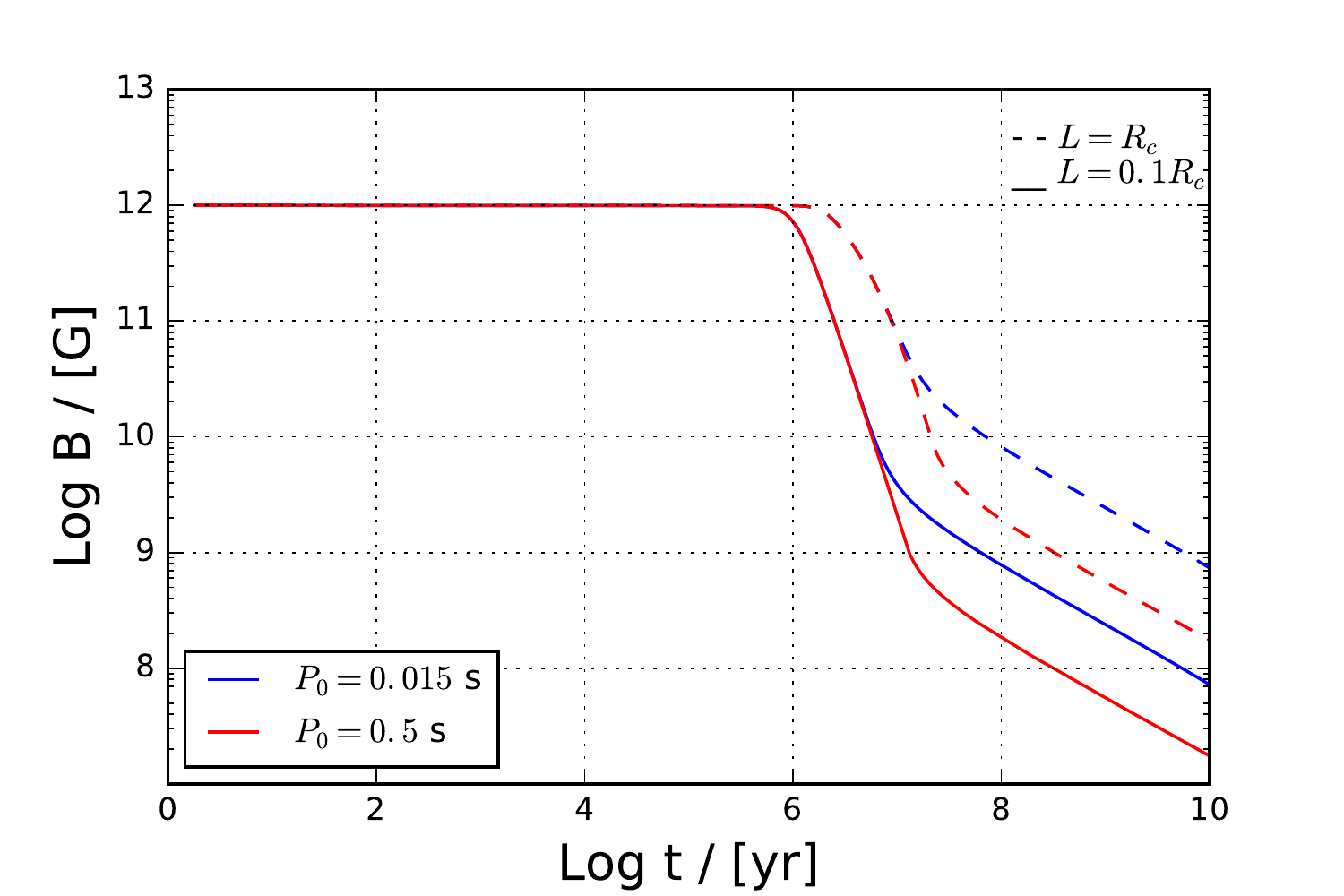} 
	\label{B-evol}
	\end{subfigure}
	
\begin{subfigure}[b]{1\textwidth} 
	\caption{Core temperature and chemical imbalance evolution}
	\includegraphics[width=1\textwidth]{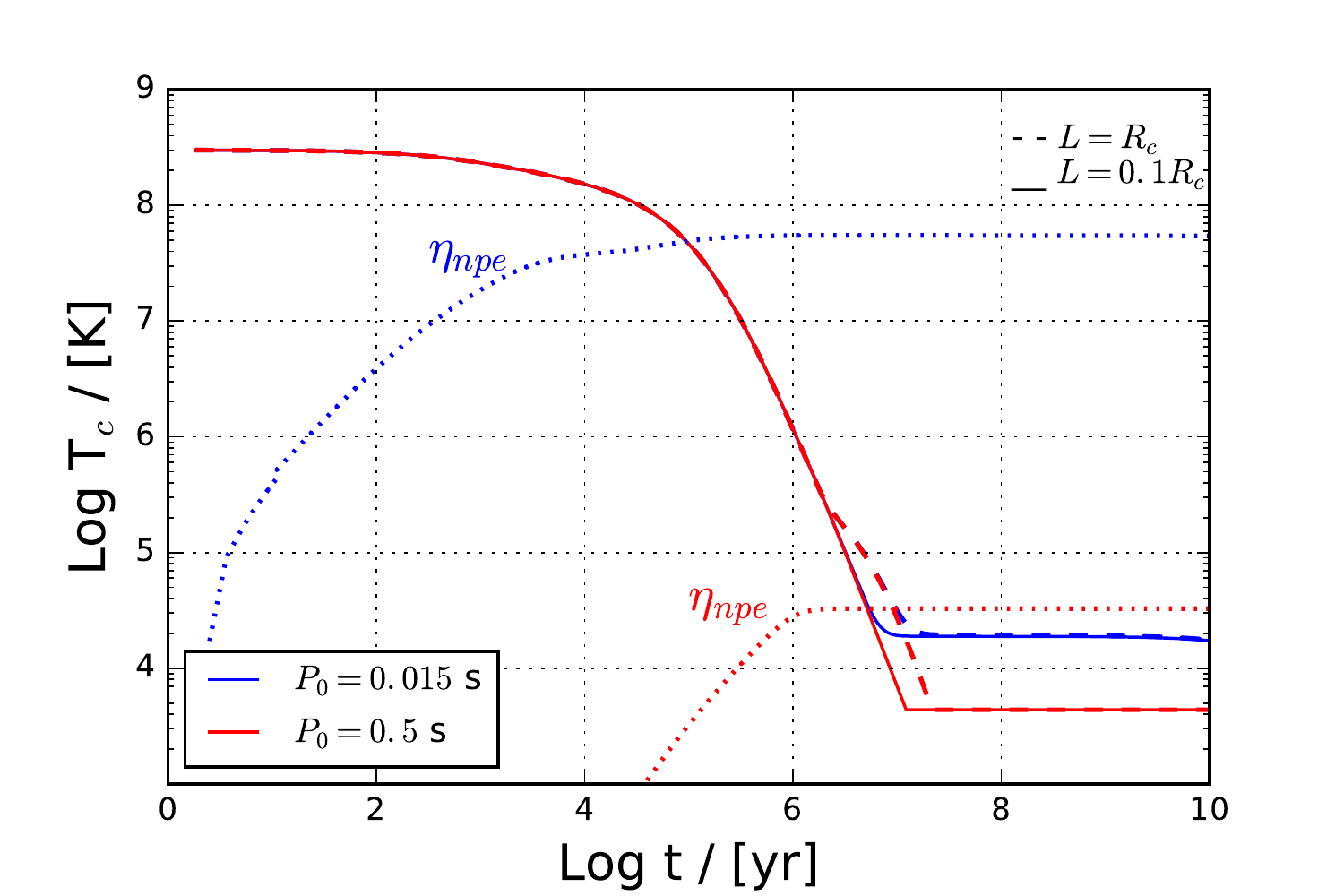}
	\label{T-evol}
	\end{subfigure}
    \vspace{-0.5cm}
\caption{Computed evolution for pulsars with initial spin periods $P_0= 0.015$ s (in blue) and $P_0= 0.5$ s (in red) and characteristic lengthscales of magnetic field gradients $L=0.1R_c$ (solid lines) and $L=R_c$ (dashed lines), with neutrino and photon cooling, rotochemical and magnetic heating, magnetic dipole spin-down, and magnetic field decay through ambipolar diffusion, assuming no accretion, superfluidity, superconductivity, or direct Urca reactions, and zero conductivity in the crust. The panels show (a) the rotation period $P$, (b) the magnetic field strength $B$, and (c) the core temperature $T_c$ and chemical imbalance $\eta_{npe}$, (divided by Boltzmann's constant $k$ in order to convert to temperature units; dotted lines), all as functions of time.}
\label{fig:evolution}
\end{figure}

In our numerical models, we consider NSs with the equation of state AV14+UVII \citep{Wiringa88} and central density $\rho_c=1.2\times 10^{15}$ g cm$^{-3}$, corresponding to a mass $M_{NS}=1.4\,M_\odot$ and radius $R_{NS}=11.4\,\mathrm{km}$, and assuming its core occupies $90\,\%$ of the total radius. Fig.~\ref{fig:evolution} shows the evolution of two hypothetical pulsars with initial magnetic field $B_0=10^{12}\,\mathrm{G}$ and with initial spin periods $P_0=15\,\mathrm{ms}$ and $0.5\,\mathrm{s}$. These periods were chosen based on PSR~J0537-6910, the fastest spinning young pulsar known to date \citep{Marshall98}, and on population synthesis models suggesting that many pulsars are born with much slower spins (e.~g., \citealt{Faucher-Giguere06}). We also explore two options for the lengthscale $L$ of the spatial variations of the magnetic field: $L=0.1R_c$ as in \citet{Goldreich92}, and $L=R_c$ as the maximum plausible value for an ordered magnetic field.\\
\\
Fig.~\ref{fig:evolution} shows that, for $t\lesssim 10^6\,\mathrm{yr}$, the magnetic field remains essentially constant (because the temperature is still high), the spin period increases as expected for a constant dipole, and the temperature is progressively reduced by passive cooling processes, for all combinations of $P_0$ and $L$ explored. For $t\gtrsim 10^6\,\mathrm{yr}$, the NS has cooled enough for the magnetic field to start decaying substantially, at a faster rater for smaller $L$. Around $t=10^7\,\mathrm{yr}$, a small bump in the temperature curve shows a reheating effect from magnetic field decay. At that time, the chemical imbalance has grown enough for rotochemical heating in the core to dominate the reheating, leading to a quasi-stationary state \citep{Reisenegger95,Fernandez05} in which the Urca reactions compensate for the contraction of the star and their heat input compensates for the photon cooling through the NS surface, keeping both $\eta_{npe}$ and $T_c$ roughly constant. Since the model NSs with faster initial rotations build up a larger imbalance, their cores will remain warmer, leading to a slower decay of the magnetic field. However, in all four models considered, the magnetic field keeps decaying roughly as $B\propto t^{-1/2}$ for all times $t\gtrsim 10^7\,\mathrm{yr}$, as expected from equation (\ref{amb_timescale}) with constant $T_c$. Therefore, the final field strength will depend on the time $t$ at which the accretion heats up the core, stopping the ambipolar diffusion.

\subsection{Evolution on the $P\dot{P}$-diagram}

\begin{figure*}
  \begin{subfigure}[b]{0.49\textwidth}
    \caption{Pulsars with He-WD companions}
    \vspace*{-0.1cm}
    \includegraphics[width=\textwidth]{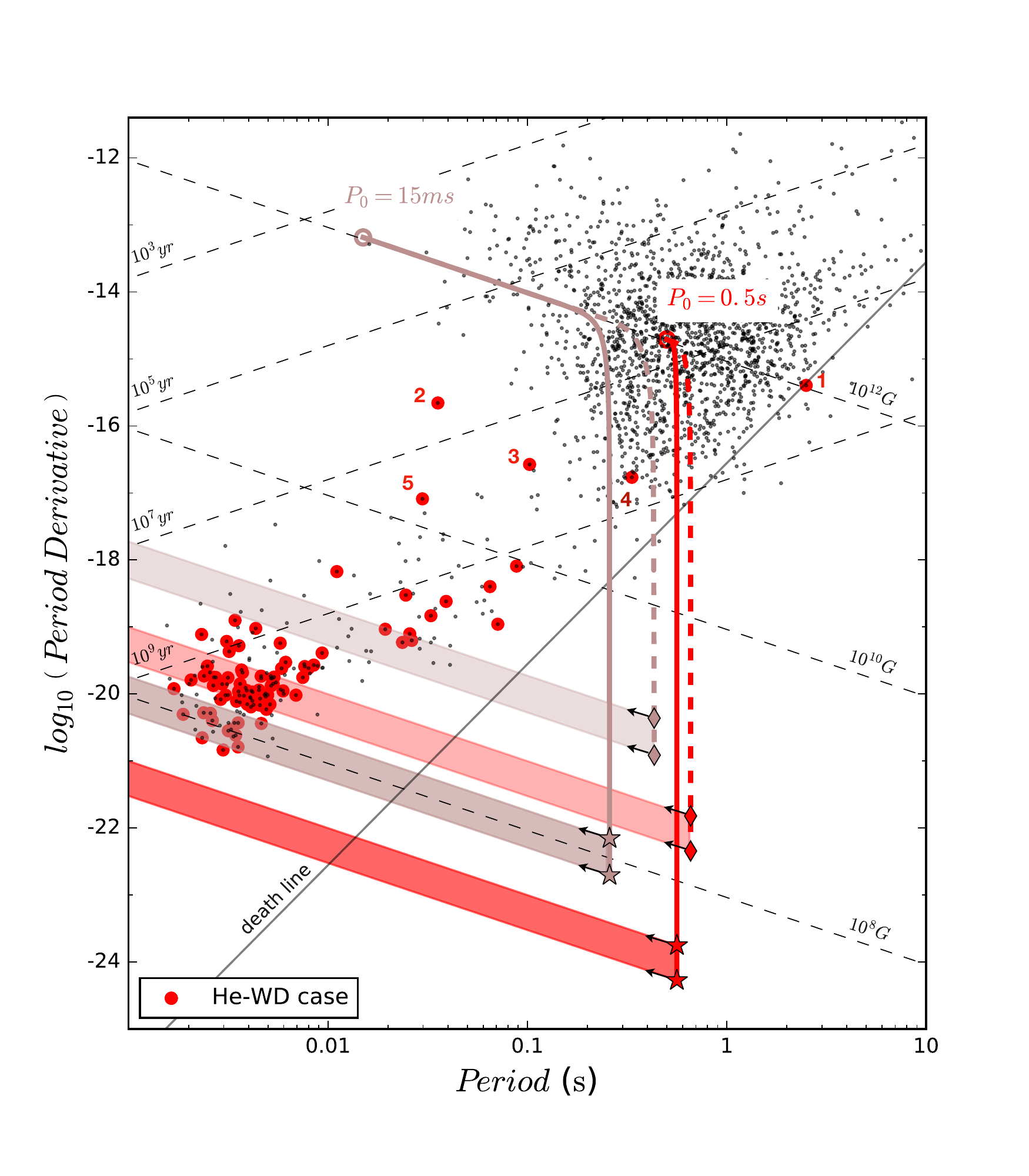}
    \label{He-case}
  \end{subfigure}
  \begin{subfigure}[b]{0.49\textwidth}
   \caption{Pulsars with CO-WD companions}
   \vspace*{-0.1cm}
    \includegraphics[width=\textwidth]{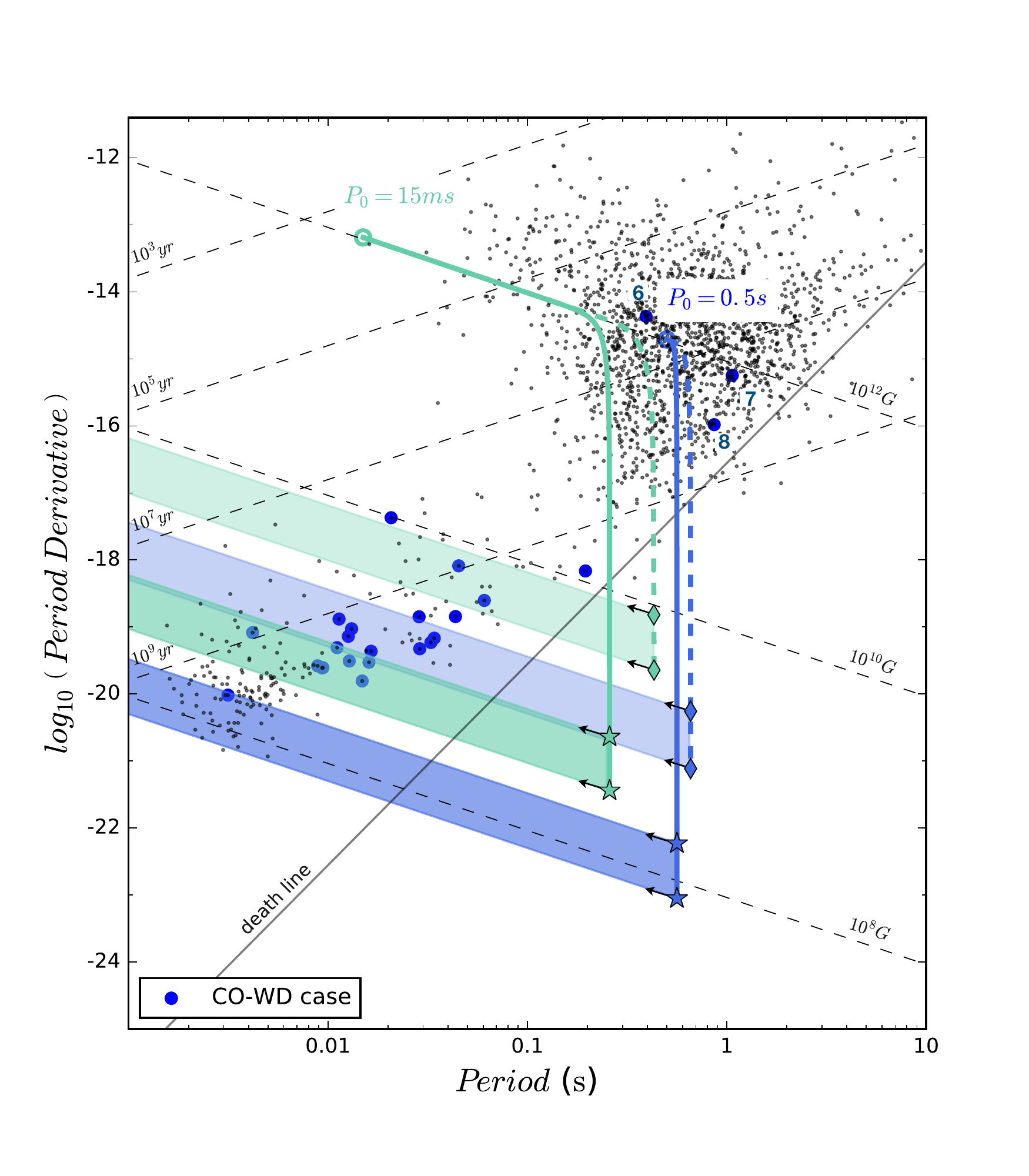}
    \label{CO-case}
  \end{subfigure}
  \begin{subfigure}[b]{0.49\textwidth}
   \vspace*{-1cm}
   \caption{Pulsars with NS companions}
    \vspace*{-0.1cm}
    \includegraphics[width=\textwidth]{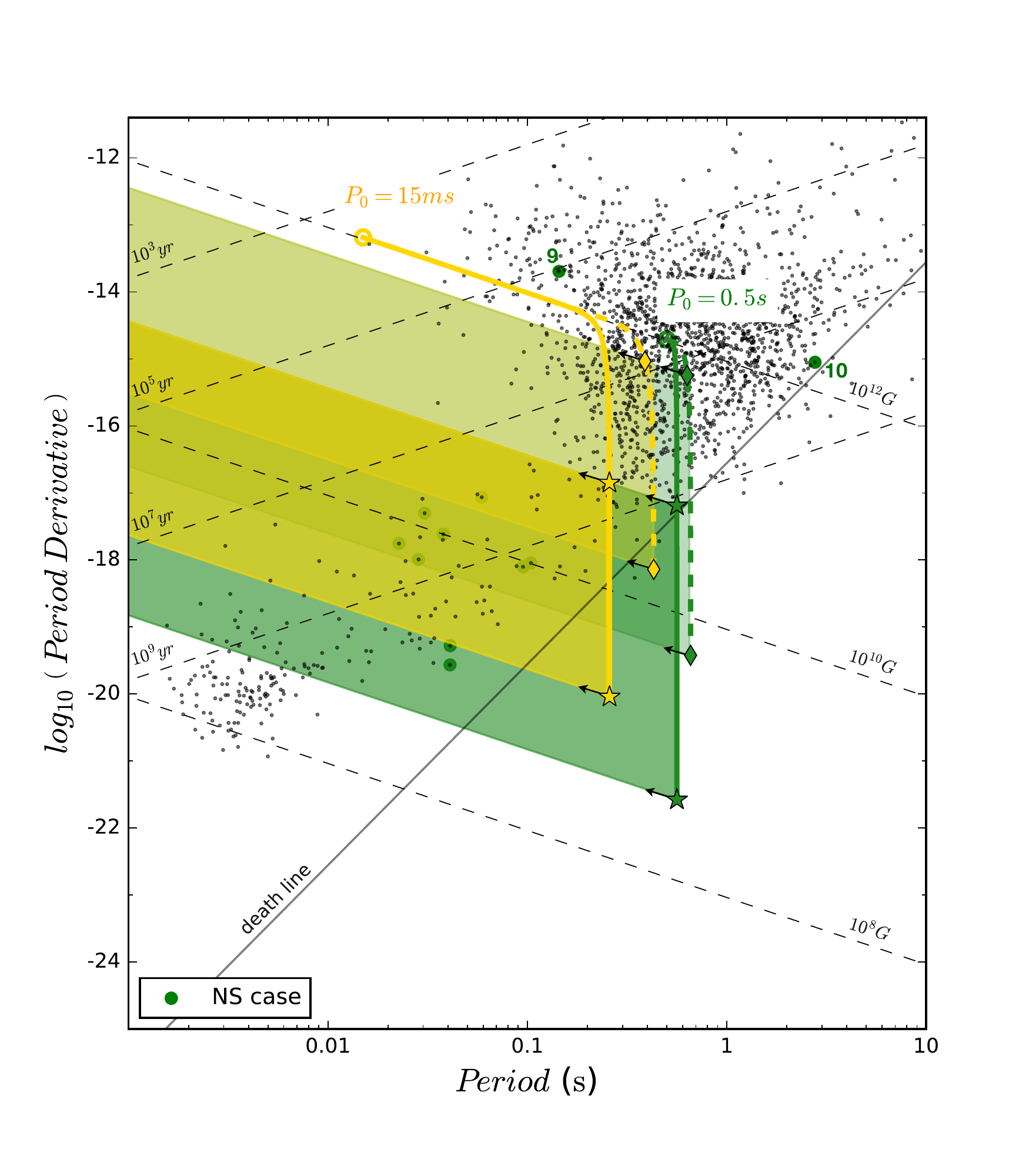}
    \label{NS-case}
  \end{subfigure}
  \vspace*{-0.8cm}
\caption{$P-\dot{P}$ diagrams for known pulsars from the \textit{ATNF pulsar catalog} (black dots), excluding objects in globular clusters. Lines of constant magnetic field strength and of constant characteristic (spin-down) age according to the magnetic dipole model are traced with grey dashed lines. In each panel, binary pulsars with a different companion type are highlighted: (a) He-WD (red circles), (b) CO-WD (blue circles), and (c) NS companions (green circles). The tracks show the evolution of a pulsar with initial spin-period of $P_0=15$ ms (in peach-color for [a], aqua-color for [b], and yellow for [c]), and another one with $P_0=0.5$ s (in red for [a], blue for [b], and green for [c]). Solid lines are used for $L=0.1R_c$, and dashed lines for $L=R_c$. The starting points of the tracks are marked with open circles and the end points with \textit{star} and \textit{diamond} symbols, based on the time available for \textit{ambipolar diffusion} ($t_{MS}$) given in Table $\ref{table_ta}$. The shaded and hatched regions represent the regions where pulsars are expected to be found once they are spun up by accretion after the magnetic field has decayed (indicated with black arrows). For all the tracks an initial magnetic field strength of $B_0=10^{12} \mathrm{G}$ was considered.}
\label{PPdots}
\end{figure*}

\begin{figure}
\vspace{-0.5cm}
\begin{subfigure}[b]{1\textwidth}
	 \includegraphics[width=1\textwidth]{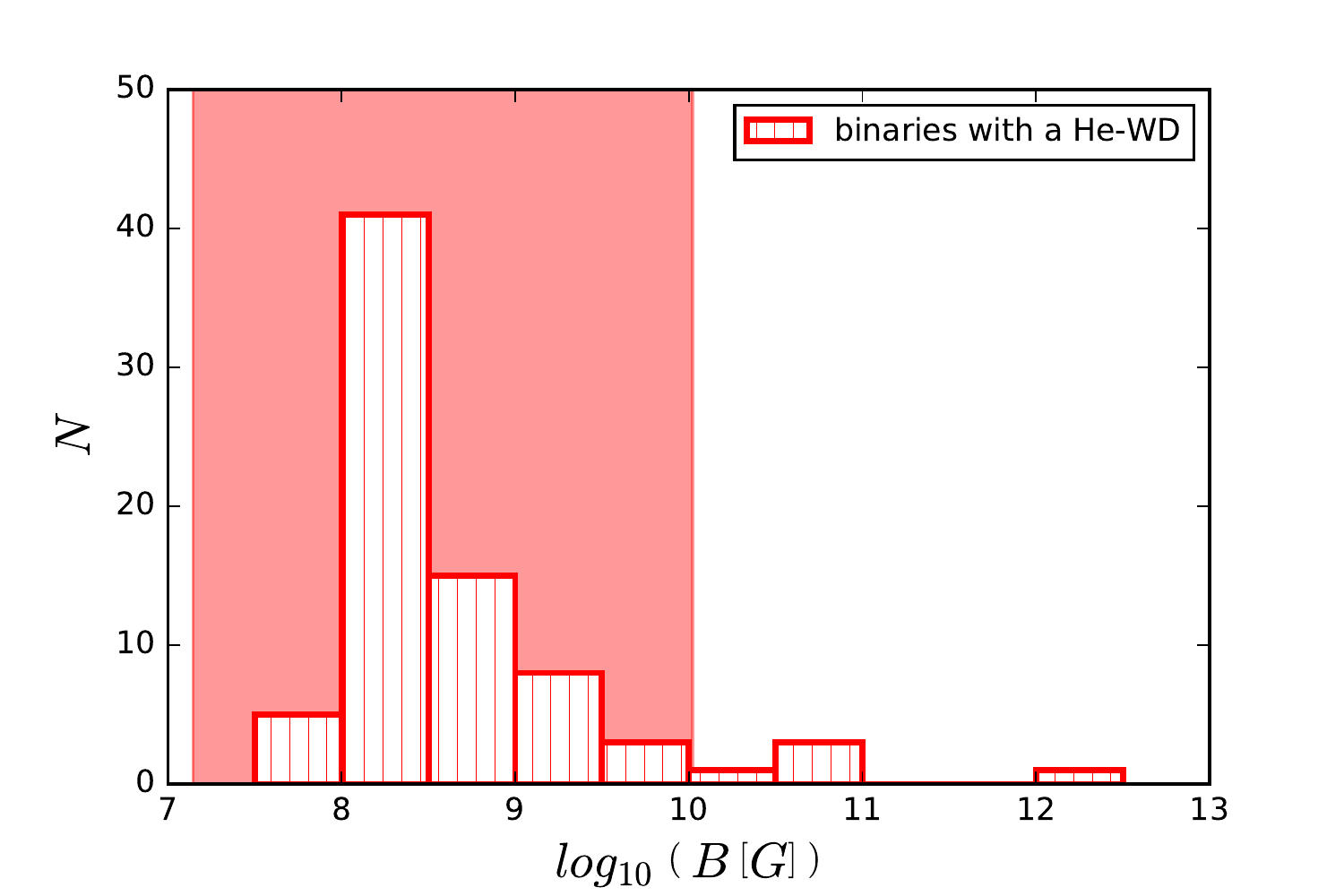} 
	\label{He-distribution}
	\end{subfigure}

\begin{subfigure}[b]{1\textwidth}
	\includegraphics[width=1\textwidth]{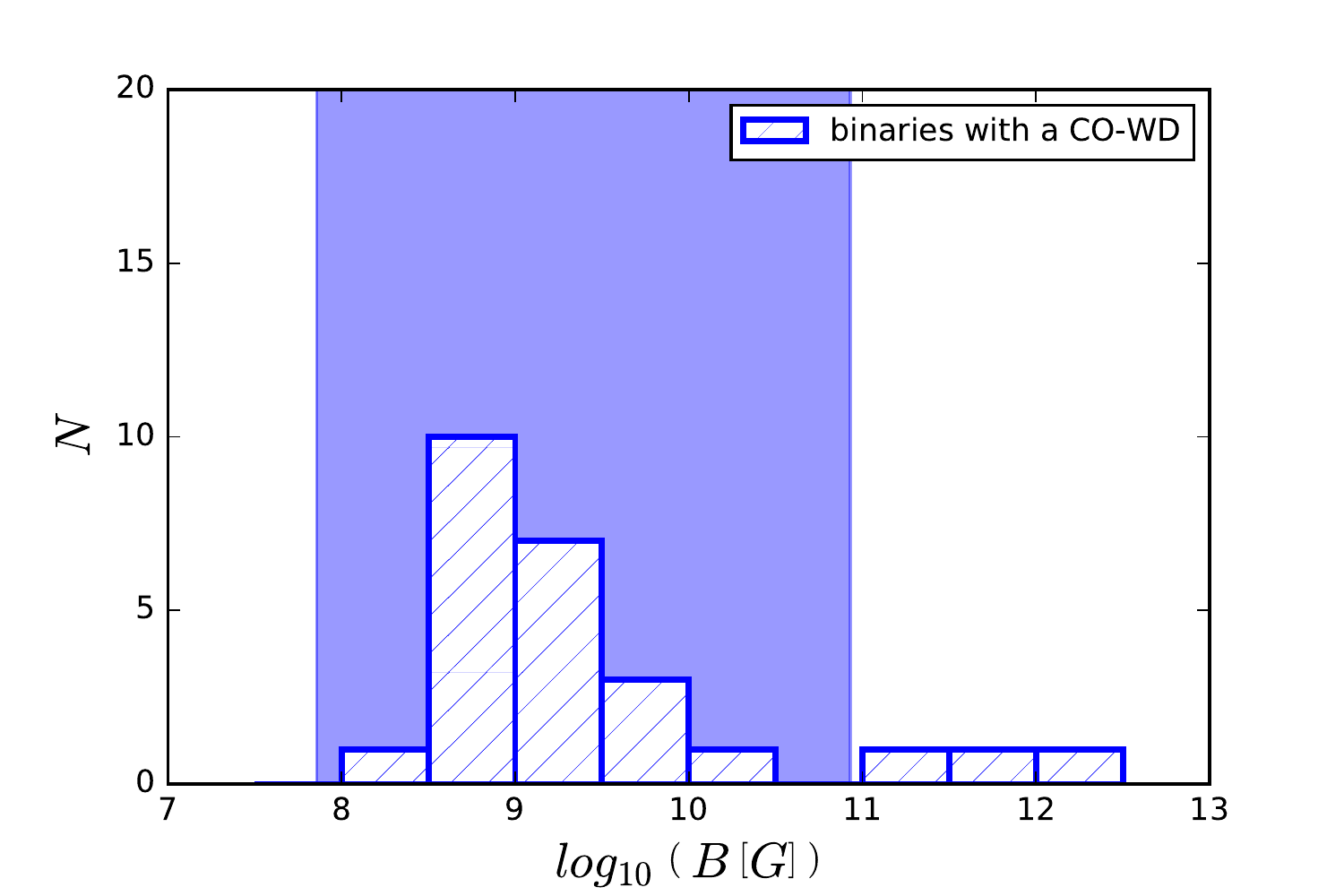} 
	\label{CO-distribution}
	\end{subfigure}
	
\begin{subfigure}[b]{1\textwidth} 
	\includegraphics[width=1\textwidth]{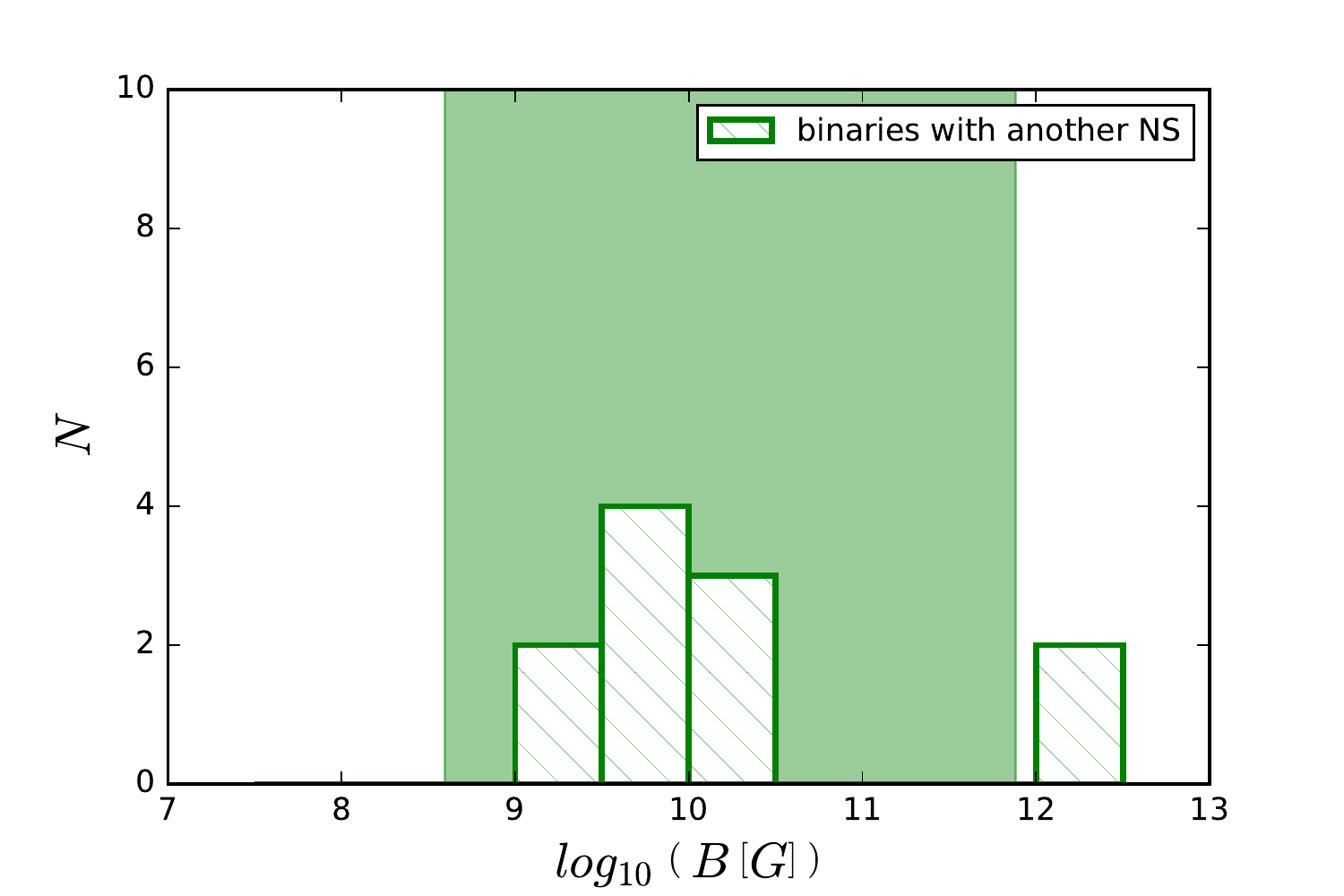}
	\label{NS-distribution}
	\end{subfigure}
    \vspace{-0.5cm}
\caption{Magnetic field strength distribution for pulsars in binary systems. Red (upper plot), blue (center plot), and green (bottom) histograms show the observed distributions of pulsars with He~WD, CO~WD or NS companions, respectively. The shaded regions of the same colors are the predictions from the ambipolar diffusion scenario for the final B domain occupied by each type of binaries by varying all the parameters: $L \in \left[0.1 R_c , R_c \right]$,  $P_0 \in \left[0.015 \mathrm{s},0.5 \mathrm{s}\right]$, $B_0 \in \left[10^{11}, 10^{14}\;{\rm G}\right]$ and the time ranges as compiled in table $\ref{prob-objects}$. }
\label{B-distribution}
\end{figure}

Since the two observables leading to the estimation of the magnetic field strength are the spin-period $P$ and its derivative $\dot{P}$ (see equation~[\ref{dipole}]), it is useful to analyze the model predictions on the so-called  $P-\dot{P}$ diagram for NSs. Each panel in Fig. $\ref{PPdots}$ is a $P-\dot{P}$ diagram containing the full sample of known pulsars in the \textit{ATNF pulsar catalogue} \citep{Manchester05}\footnote{Version 1.54; http://www.atnf.csiro.au/people/pulsar/psrcat.}, except for objects located in globular clusters, where the dense enviroment can cause companion exchange and hence the current companion may not be suitable to infer the pulsar evolution history. Each panel highlights binary pulsars with a different companion type, showing that pulsars with less massive companions also tend to have weaker magnetic fields.\\
\\
It can be seen in Fig. $\ref{PPdots}$ that the evolution from classical pulsars to MSPs in the ambipolar diffusion scenario can be clearly divided into three successive stages:
\begin{itemize}
\item[(i)] The pulsar spins down (and cools) at roughly constant magnetic field. 
\item[(ii)] The magnetic field decays through ambipolar diffusion, while the rotation period approaches a constant value. During this stage, the NS crosses the ``death line'', at which the pulsar activity ceases. Since at this point the temperature is also extremely low, the NS should become undetectable. 
\item[(iii)] Accretion from the binary companion increases the temperature, choking ambipolar diffusion and thus the field decay, and accelerates the rotation, moving the pulsar back across the death line into the MSP region of short spin periods and weak magnetic fields.
\end{itemize}
Clearly, the different times available for magnetic field decay, depending on the companion mass, produce a trend in the final magnetic fields that roughly follows that in the observed systems. This can also be seen in Fig. $\ref{B-distribution}$. Our simplified model does particularly well in reproducing the ranges of magnetic field strengths in which the bulk of pulsars of each companion class are found, although there is a number of outliers, specifically towards stronger fields than predicted, which we discuss below.\\

\begin{figure}
\begin{center}
\hspace*{-1cm}
\includegraphics[width=1.2\textwidth]{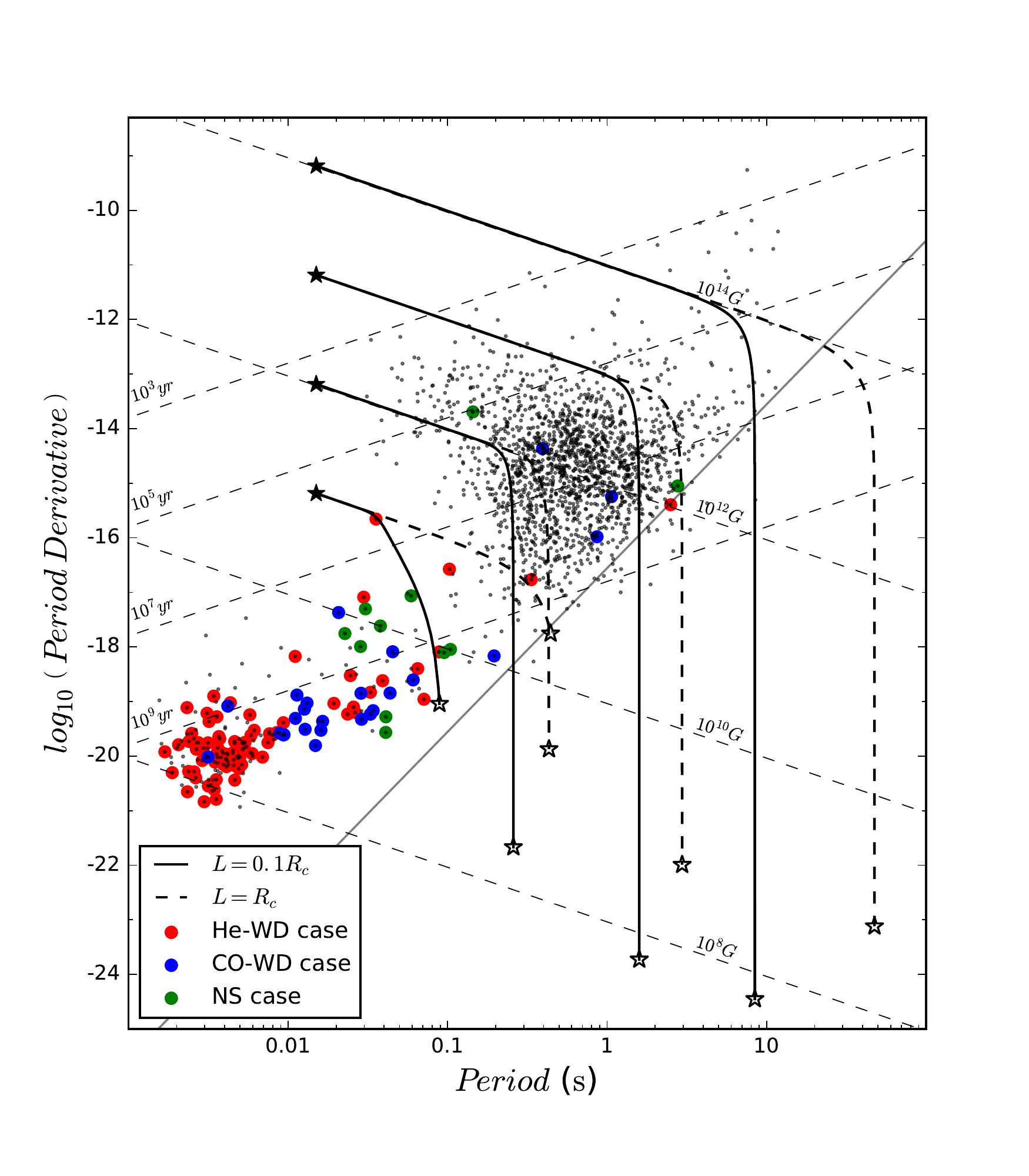} 
\caption{Tracks followed by a pulsar born with $P_0=15\,\mathrm{ms}$ for different initial magnetic field strengths $B_0=10^{11}, 10^{12}, 10^{13}$, and $10^{14}\,\mathrm{G}$, considering $L=0.1 R_c$ (solid lines) and $L=R_c$ (dashed lines). The starting point in each track is tagged with a filled star, while the final point reached after $10^9\,\mathrm{yr}$ is marked with an open star. Binary pulsars with a He~WD companion are highlighted with red dots, pulsars with a CO~WD with blue dots, and pulsars with a NS companion with green dots.}
\label{B-tracks}
\end{center}
\end{figure}
In Fig. $\ref{B-tracks}$, we explore the dependence of the evolution on the initial magnetic field $B_0$ of the pulsar. For large values of $B_0$, the pulsar spins down at roughly constant $B\approx B_0$ for $\sim 10^6\,\mathrm{yr}$, reaching longer periods for stronger $B_0$. At this point, the star has cooled enough for ambipolar diffusion to set in, rapidly reducing the magnetic field and thus essentially stopping the spin-down. Contrary to the final rotation period, the final value of $B$ is essentially independent of $B_0$. On the other hand, for small $B_0$ and $L=R_c$, the ambipolar diffusion time remains long even after $t\sim 10^7\,\mathrm{yr}$, when the temperature has been strongly reduced. Therefore the spin-down continues until much later, and only then the ambipolar diffusion sets in and reduces the field. 

\section{Discussion}
\label{sec:Discussion}

\subsection{Outliers}

Fig. $\ref{B-distribution}$ shows that the actual distribution of pulsar magnetic fields for each companion class has a peak at about the expected location, but also a tail towards stronger magnetic fields that cannot be explained by a straightforward application of our model. Table $\ref{prob-objects}$ lists the problematic objects, all of which are located in the region of classical (isolated, non-recycled) pulsars with relatively strong magnetic fields ($B\sim 10^{10-12}\,\mathrm{G}$) rather than that of ``recycled'', weak magnetic field MSPs ($B\sim 10^{8-9}\mathrm{G}$), as expected at least for those with WD companions (objects 1-8). Since we are not considering pulsars in globular clusters, it is very unlikely that any of these Galactic field pulsars would have exchanged their companion during their lifetime.\\
\\
Disregarding object 2 (where no eccentricity has been measured, nor an upper limit estimated) and 5, a common characteristic of the problematic objects is the relatively large eccentricity of their orbits, which contradicts the highly circular ($e\la 10^{-4}$) orbits expected for binary systems following mass transfer and tidal interactions \citep{phi92,sfa+19}. In addition, objects 1, 2, 3, 4 and 8 are wide binaries with orbital periods longer than 200 days. These systems are therefore expected to have accreted less material due to their shorter X-ray lifetime \citep{Tauris99,prp02}, as is also reflected in their relatively long spin periods. All fully recycled ($<10\;{\rm ms}$) pulsars have orbital periods below 200~days \citep{Tauris12}. The above characteristics are therefore well supported by the standard scenario involving accretion-induced B-field decay.\\
\\
The feature regarding relatively significant eccentricity and high magnetic fields in pulsars with WD companions not only challenges the ambipolar diffusion scenario, where the weak MSP fields are a consequence of the old age of the NS, but also the standard accretion scenario for object~5 (PSR~J1841+0130) which has an orbital period of only 10~days. In this system, if the NS was formed before its WD companion, the mass transfer should not only have reduced the magnetic field strength, but also have circularized the orbit \citep{Verbunt95}. NSs born after the WD companion \citep{Dewey87}, accretion-induced collapse (AIC) of a massive WD \citep{Taam86}, and a merger of two WDs \citep{Saio85} in a triple system are among the scenarios suggested as alternatives channels for the formation of binary unusual NSs.\\
\\
\cite{Tauris00} used population synthesis to analyze the evolution of interacting binaries leading to the formation of systems with a young NS member orbiting an old WD, i.e. systems where the NS was formed after the WD due to mass reversal resulting from mass transfer between the progenitor stars. It was suggested that PSR~J1141$-$6545 and PSR~B2303+46 (objects~6 and 7, respectively, in Table~\ref{prob-objects}) are systems where the pulsar was formed after the WD. From the standard scenario, the high magnetic fields observed here is a natural consequence of the NS being non-recycled. From the ambipolar diffusion scenario, the high magnetic fields observed would be a consequence of the young age of the NS. The high magnetic field and large eccentricity in PSR~J1822$-$0848 (object~1) and PSR~B0820+02 (object~8) may suggest these pulsars belong to the same population. Nevertheless, \cite{Tauris12} argued that these eccentricities are not large enough for such wide orbits based on expectations for an unperturbed imprint of the supernova explosion if the NS formed after the WD. Instead, it was proposed that these pulsar binaries originate from very wide-orbit LMXBs resulting in CO remnants \citep{Tauris99}, and that PSR1822$-$0848 experienced a mild spiral-in from an almost unbound common envelope (CE).\\
\\
In the AIC scenario, accretion onto a massive WD ($1.1-1.3 M_{\odot}$; \citealt{Tauris13} and references therein) may lead to the formation of a super-Chandrasekhar WD, which can collapse immediately or after it loses sufficient spin angular momentum after the accretion phase \citep{Freire14}. As found by \citealt{Tauris13}, binary pulsars formed via AIC and which have He~WD companions are expected to have orbital periods of either $10-60\;{\rm days}$ or $>500\;{\rm days}$, depending on whether these systems originated from a main-sequence or a giant-star donor. The exact orbital period boundaries are uncertain and depend on details of WD accretion physics.\\
\\
For standard evolution of an X-ray binary with a NS accretor, when the donor is a giant star, the outcome is a mildly recycled NS (due to the short duration of the RLO) with a He~WD or CO~WD companion -- depending on the initial orbital period and the mass of the donor -- with orbital periods spanning between hundreds and more than a thousand days. We propose that PSR~J1840$-$0643, PSR~J1711$-$4322 and PSR~J1803$-$2712 (objects~2, 3, and 4 respectively) might be formed in a similar way as PSR~B0820+02 (i.e. RLO in a wide-orbit LMXB), with the difference that the initial orbital period in these systems were slightly smaller and therefore ended up as a He~WD systems, given the WD mass--orbital period correlation \citep{Tauris99}. If, however, magnetic flux conservation during the AIC process applies, magnetic field strengths above $10^{10}\;{\rm G}$ (as observed in all objects in Table~\ref{prob-objects}) can be expected for the formed NS if we consider the broad range in magnetic field strengths covered by magnetic WDs. (We note that only $\sim 10^4\;{\rm G}$ is needed for a WD progenitor to produce a post-AIC field of $10^{10}\;{\rm G}$.) 
Furthermore, post-AIC systems are expected to undergo additional mass transfer after formation of the NS \citep{Tauris13}.
Magnetic field strengths in WDs can be as high as $10^9\;{\rm G}$, as revealed from Zeeman and cyclotron effect \citep{Ferrario15}.
If such WDs collapse due to AIC they might produce NS with magnetar-like magnetic fields.\\
\\
Another alternative channel for the formation of NSs is through the merger of two close massive WDs. In this scenario, the evolution of two intermediate-mass close stars may lead to the formation of a double WD system in which orbital energy is released via gravitational wave radiation, leading to a  merger forming the NS \citep{Saio85}. In order to connect this formation channel with the current observed pulsars with a WD companion, a third member in a wide orbit needs to be present.\\
\\
\citet{Rappaport13} estimated that at least 20$\%$ of the close binaries in the Galactic disk contain a third member in a wide orbit. In the WD merger scenario, a high eccentricity is expected due to the release of binding energy, if the binary survives the effect of the merger. It can be speculated that this scenario might explain the high eccentricity of 0.058 seen for PSR~J1822$-$0848 in a binary with a He~WD, although RLO from the tertiary star to the inner binary after the WD merger event would probably have circularized the orbit. The high magnetic field strength observed might be due to the young age of this pulsar or be a product of a relatively large seed magnetic field in the WDs.\\
\\
Weak spiral-in due to an unbound common envelope of an asymptotic giant branch star has been claimed in order to explain pulsars with high magnetic fields in wide, eccentric binaries with $P_b < 1000$ days like PSR J1822-0848 \citep{Tauris12}. This post-MS phase where the RLO was not efficient would explain the almost non-recycled characteristics of some pulsars in the accretion scenario, but not in the ambipolar diffusion scenario, where magnetic field decay is still expected due to the long MS timescale of several Gyr for the WD progenitor -- unless the NS formed somewhat recently through some of the alternative scenarios discussed above.\\ 
\\
PSR~J0737$-$3039B is the slower pulsar in the ``double pulsar'' binary, and thus the second NS to form in the system \citep{Lyne04}. The same may be the case for PSR~J1906+0746, in which the companion star is not detected and its properties are therefore unknown (\citealt{Lorimer06}; \citealt{Leeuwen15}).

\begin{table*} 
 \caption{Summary of properties of the outlier objects labeled in Fig. $\ref{PPdots}$, from the \textit{ATNF pulsar catalog} (\citealt{Manchester05}; http://www.atnf.csiro.au/people/pulsar/psrcat). In the last column, we add our proposed formation scenario for each system, which might make it consistent with our model. Abbreviations: LMXB (Low mass x-ray binaries), CE (common-envelope phase), AIC (accretion-induced collapse), WDNS (white dwarf born before the neutron star) and 2nd NS (second neutron star to be formed).
\label{prob-objects}}
 \begin{tabular}{lccccccc}
 \hline
Companion  & Object & Pulsar & Spin period & Magnetic field   & Orbital period & Eccentricity & Proposed\\
type & number & name & $P\,[\mathrm{s}]$ & $B\,[\mathrm{G}]$ & $P_b$ [days] & & formation\\
  \hline
  \hline
He~WD & 1 & J1822-0848 & 2.5045 
& $1\times 10^{12}$ & 286 & 0.058 & wide-LMXB and CE \\
      & 2 & J1840-0643 &     0.0355
      & $9\times 10^{10}$ & 937 & Unknown & wide-LMXB or AIC \\
      & 3 & J1711-4322 &   0.1026
      & $5\times 10^{10}$ & 922 & 0.002 & wide-LMXB or AIC \\
      & 4 & J1803-2712 &      0.3344 
      & $8\times 10^{10}$ & 407 & 0.005 & wide-LMXB or AIC \\
      & 5 & J1841+0130 &  0.0297 
      & $2\times 10^{10}$ & 10.5 & $8\times 10^{-5}$ & AIC \\
\hline
CO~WD & 6 & J1141-6545 &   0.3938 
& $1\times 10^{12}$ & 0.197 & 0.171 & WDNS \\
      & 7 & B2303+46   &  1.0663 
      & $8\times 10^{11}$ & 12.3  & 0.658 & WDNS \\
      & 8 & B0820+02   &  0.8648 
      & $3\times 10^{11}$ & 1232  & 0.011 & wide-LMXB \\
\hline 
NS    & 9 & J1906+0746 &  0.1440 
      & $2\times 10^{12}$ & 0.165 & 0.085 & 2nd NS? \\
     & 10 & J0737-3039B &  2.7734 
     & $2\times 10^{12}$ & 0.102 & 0.087 & 2nd NS \\
  \hline
\end{tabular}

\end{table*}
\subsection{Crustal resistivity and impurity parameter}

If the outlier objects do not have a different origin, but their NS component was formed before its WD companion, we would need to re-examine our assumptions. A possible explanation for the mismatch between the magnetic field strength predicted in the diffusion scenario
and the observed magnetic field strength of pulsars in wide binaries lies in one of the strongest assumptions of the diffusion model, namely that the Ohmic dissipation of currents in the crust (equation \ref{Ohm_timescale}) is much faster than the ambipolar diffusion in the cool ($T_c\sim 10^4\,\mathrm{K}$) core of an old, non-accreting NS, so the latter process controls the decay of the magnetic field. This requires the crust before accretion to have a moderately high impurity parameter, $Q\gtrsim 1$. If this is not the case, the magnetic field will remain ``frozen'' into the crust and will not decay on the ambipolar diffusion time.\\
\\
Regardless of the actual value of $Q$, we do not expect strong variations from one pulsar to another, unless it is modified by accretion. If $Q$ is very small in all pristine pulsars, our model would not be viable and some version of the accretion scenario would be favored.

\subsection{Shortcomings of our model}

We emphasize once more that the model we use is a very simplified toy model. 

First, instead of considering a three-dimensional vector field and the particle density perturbations caused by it (e.~g., \citealt{Goldreich92,Hoyos08,Castillo17}), we characterized the magnetic field strength by a single number $B(t)$, whose time-variation is controlled by a characteristic time $t_{AD}$ that depends on a constant (and somewhat arbitrary) characteristic length scale $L$. Current two-dimensional (axially symmetric) simulations \citep{Castillo17} show that the magnetic field might reach an equilibrium configuration in which ambipolar diffusion does not lead to further evolution. On the other hand, such equilibria might be subject to three-dimensional instabilities \citep{Mitchell15}, which have not yet been explored in the two-fluid model for a neutron star core.

Also, the microphysics going into the estimate of $t_{AD}$ is strongly simplified. As in previous work \citep{Goldreich92,Hoyos08,Castillo17}, we are assuming that the neutron star core is composed exclusively of non-superfluid neutrons, non-superconducting protons, and electrons (without any ``exotic'' particles such as muons or hyperons) and cools through modified Urca processes. At the very low temperatures at which the evolution occurs in this model, it is likely that at least some part of the neutron star core will be in a superfluid or superconducting state, which would affect the dynamics (particularly $t_{AD}$), and the heat capacity and the neutrino cooling and heating rates (e.~g., \citealt{Petrovich10}). However, it is not clear at present how much of the core is affected, what values are taken by the energy gaps, and whether the protons form a type-I or type-II superconductor. Furthermore, in spite of recent progress \citep{Gusakov17,Passamonti17,Kantor18,Drummond18}, there is not yet a self-consistent model for the dynamics of this kind of matter in the presence of rotation (which produces quantized vortices in the neutron superfluid) and a magnetic field (which is confined into quantized flux tubes in a type-II superconductor and into more complex domains if the superconductor is of type-I). 

Given these large uncertainties, it is perhaps surprising that our very simple model produces a semi-quantitative agreement with the available data, which we take to indicate that a better model might also give a good (hopefully better) description. Of course, given the many simplifications, the agreement might just be a coincidence, and a more realistic model might not fit the data well.


\section{Conclusions}
\label{sec:Conclusions}
We explored a scenario in which the weak magnetic fields of MSPs are caused by ambipolar diffusion in the NS core in the non-superfluid/superconductor regime. This process is effective while the core is cool, i.e. in the time interval after the initial cooling of the NS and until its reheating by mass transfer from its companion star. The duration of this time interval is approximately set by the main-sequence lifetime of the companion star, since it most often fills its Roche~lobe once it becomes a giant. For the core to drive the magnetic field decay, the currents in the crust must be dissipated by ohmic diffusion before the temperature drops and ambipolar diffusion starts playing a role. The previous condition ($t_{Ohm}\ll t_{AD}$) holds if the \textit{impurity parameter Q} is high enough for the impurity scattering to dominate and to suppress the flow of currents in the crust. In this scenario, the crust behaves as an extension of the near-vacuum outside the star.\\
\\
With a simple model, we can roughly reproduce the magnetic field strengths of the bulk of the pulsars in binary systems with He~WDs, CO~WDs, and NSs, given the main-sequence lifetime inferred for the progenitor of the respective companion. There are, however, a certain number of outliers that have substantially stronger magnetic fields than predicted by straightforward application of this model (and which are also problematic for the more standard, accretion-induced field decay scenario), and most of which have relatively wide orbits and moderate to large eccentricities. A possible explanation is that these belong to a different population in which the NS was formed relatively recently, through alternative channels such as: WD--NS formation reversal, accretion-induced collapse of a WD, or merger of the inner two WDs in a triple system. Otherwise, one would have to invoke a very pure crust with a high conductivity, which would not allow the magnetic field to decay unless it is somehow driven by accretion.\\
\\
We motivate further analysis of the formation channels leading to wide, eccentric binary systems containing pulsars whose magnetic field has not decayed significantly, as they challenge our understanding of pulsar evolution and particularly how the recycling process to form a MSP proceeds.

\section*{Acknowledgements}
We are grateful to C. Espinoza, J. Tan and P. Freire for useful conversations about pulsar evolution. M. Cruces and A. Reisenegger acknowledge funding from FONDECYT Regular Projects 1150411 and 1171421, CONICYT Chile-Germany International Cooperation Grant DFG-06, and the Center for Astronomy \& Associated Technologies (CATA; CONICYT project Basal AFB-170002). This work made extensive use of the SAO/NASA Astrophysics Data System (ADS; {adsabs.harvard.edu}).
\addcontentsline{toc}{section}{Acknowledgements}



\bibliographystyle{mnras}

\bibliography{main}

\bsp	
\label{lastpage}
\end{document}